\documentclass{nsr}

\usepackage{amsmath,graphicx,array}
\usepackage{dcolumn,soul}%

\usepackage{amsthm}
\usepackage[figuresright]{rotating}%
\usepackage{algorithm, algorithmicx, algpseudocode}
\usepackage{listings}%
\usepackage{hyperref}
\usepackage{multirow, threeparttable}
\usepackage{pdfpages}

\makeatletter
\def\uns{\ifmmode\,\else$\,$\fi}%

\makeatother

\jvol{XX}
\jnum{X}
\jyear{Year}
\doi{10.1093/nsr/XXXX}
\received{XX XX Year}
\revised{XX XX Year}
\accepted{XX XX Year}

\begin{document}

\dhead{RESEARCH ARTICLE}

\subhead{PHYSICS}

\title{Ultrahigh-energy gamma-ray emission associated with black hole-jet systems}

\author{The LHAASO Collaboration$^{\dagger}$,*}



\authornote{\textbf{Corresponding authors.} Emails: ryliu@nju.edu.cn; jianli@ustc.edu.cn; k-wang@smail.nju.edu.cn; yangrz@ustc.edu.cn; yuyh@ustc.edu.cn; chensz@ihep.ac.cn; hushicong@ihep.ac.cn}
\authornote{The LHAASO Collaboration
authors and affiliations are listed in the end of the paper and also Supplementary Materials, which is available with this article at the \em{National Science Review} website.}

\abstract[ABSTRACT]{Black holes (BH), one of the most intriguing objects in the universe, can manifest themselves through electromagnetic radiation initiated by the accretion flow. Some stellar-mass BHs drive relativistic jets when accreting matter from their companion stars, forming microquasars. Non-thermal emission from the radio to tera-electronvolt (TeV) gamma-ray band has been observed from microquasars, indicating the acceleration of relativistic particles.
{ Here we report detection of four microquasars (SS~433, V4641~Sgr, GRS~1915+105, MAXI~J1820+070) of spectrum extending to the ultrahigh-energy (UHE; photon energy $E>100$\,TeV) band and one microquasar (Cygnus X-1) of spectrum approaching 100\,TeV, using the Large High Altitude Air Shower Observatory (LHAASO). Notably, the total emission associated with SS~433 cannot be interpreted with a single leptonic component.} In the UHE band, its emission is in spatial coincidence with a giant atomic cloud,  which is consistent with a hadronic origin. An elongated source is discovered from V4641~Sgr with the spectrum continuing up to 800\,TeV.
The detection of UHE gamma rays demonstrates that accreting BHs and their environments can operate as extremely efficient accelerators of particles out of 1\,peta-electronvolt (PeV), suggesting microquasars to be important contributors to Galactic cosmic rays especially around the `knee' region.}



\maketitle

\section{INTRODUCTION}\label{sec1}

The accretion process of BHs may lead to various astrophysical phenomena over a wide range of radiation bands through converting the gravitational energy of falling matters or rotational energy of BHs into energetic particles. Presence of powerful jets is tightly correlated with the accretion state and may usually be an indication of high accretion rate and efficient extraction of BHs' rotational energy\cite{BZ77, Fender06}. Accreting BHs have long been suggested as efficient {accelerators of particles exceeding 100\,TeV}\cite{Aharonian98_jet, Sudoh20, Khangulyan24}, but it is not clear how much is their contribution to the measured cosmic ray (CR) flux. Gamma-ray emission is widely considered as an effective probe of energetic particle acceleration processes\cite{Bosch-Ramon09}. While UHE photons from activities of accreting supermassive BHs in the distant universe (i.e., active galactic nuclei, AGN) cannot be detected due to severe absorption by the extragalactic background light and the cosmic microwave background (CMB) permeating the universe, Galactic microquasars\cite{Mirabel99_review, Remillard06} provide us with an opportunity to look closely into the acceleration of high-energy particles in BH-jet systems.

Among dozens of BH-jet systems identified in the Milky Way thus far, only a few of them have been detected in high-energy (HE; energy above 0.1\,GeV) and very-high-energy (VHE; energy above 0.1\,TeV) gamma-ray bands: previous observations detected sporadic HE gamma-ray emission\cite{Sabatini10, Malyshev13, Zanin16} and {hints of} hour-scale TeV flares\cite{MAGIC07_X1} from Cygnus X-1; Individual HE flares were reported from V404 Cygni\cite{Loh16, Piano17, Xing21} although the validity of these results was challenged by a later independent analysis\cite{Harvey21}; A  hint of a VHE flare was reported from GRS~1915+105\cite{HEGRA98} { and a possible GeV gamma-ray counterpart of the microquasar was discovered recently\cite{Marti-Devesa2025}}. So far, two microquasars, SS~433\cite{bordas15,HAWC18_SS433,sun19, HESS24_SS433} and V4641~Sgr\cite{HAWC24}, have been unambiguously detected up to several tens of TeV and to 200\,TeV respectively, implying acceleration of $\sim 100\,$TeV and $\sim 1\,$PeV particles.

\section{LHAASO OBSERVATIONS}\label{sec2}

The Large High Altitude Air Shower Observatory (LHAASO) is a  km$^2$-scale instrument for the gamma-ray detection over a wide energy range from 1 TeV to a few PeV. As of now, LHAASO has reached an exceptional sensitivity of $10^{-14}\rm \, erg~cm^{-2}s^{-1}$ at approximately 100\,TeV {photon energy} for point-like sources, making it an ideal detector for UHE gamma-ray sources that probe accelerators of CRs around and beyond PeV energies.
In the field of view of LHAASO, there are {  10 transient} microquasars with dynamic evidence for BHs as central engines \cite{BlackCAT2016} (see Table~1).
For persistent microquasars, Cyg X-1 is confirmed to hold a BH as compact object\cite{Cyg_X1_2021Sci} and SS 433 is very likely hosting a BH\cite{SS4332021MNRAS.507L..19C} and thus included in this paper.
{Cyg X-3 is another good BH candidate \cite{CygX-3_2013MNRAS.429L.104Z} but will be reported in a dedicated LHAASO study.}
{Since these objects are located at a distance of no more than 10\,kpc from Earth, where the absorption of UHE photons is not very strong, LHAASO is able to carry out detailed studies.}

\begin{figure*}[htbp]
\centering
\includegraphics[width=1.0\textwidth]{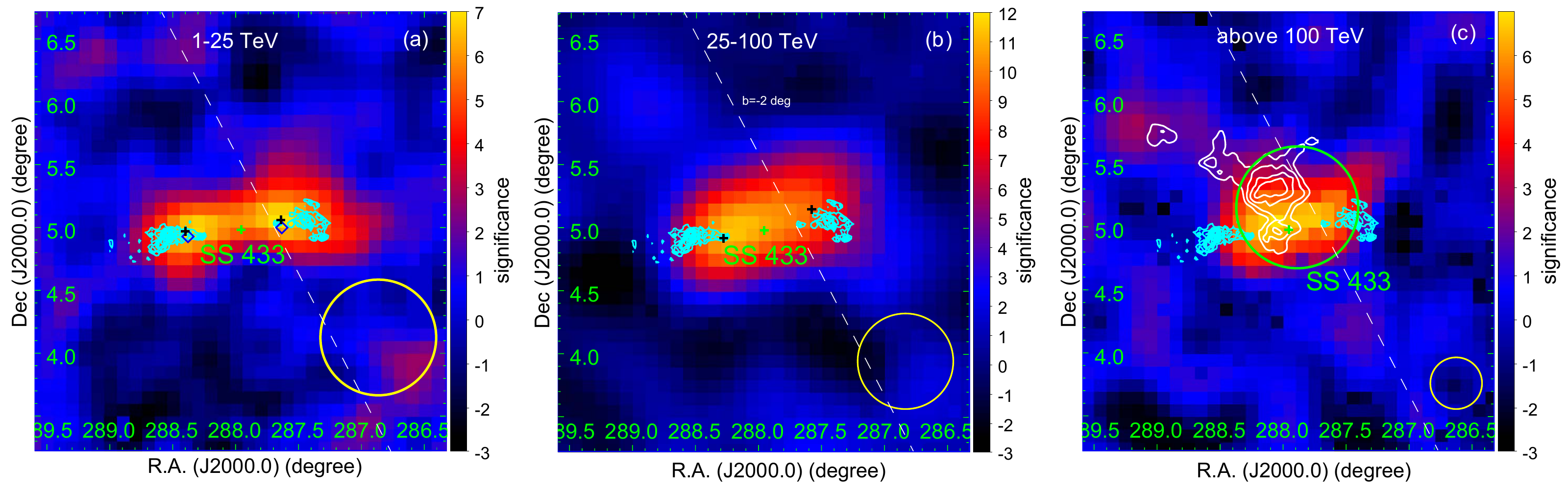}
\includegraphics[width=1\textwidth]{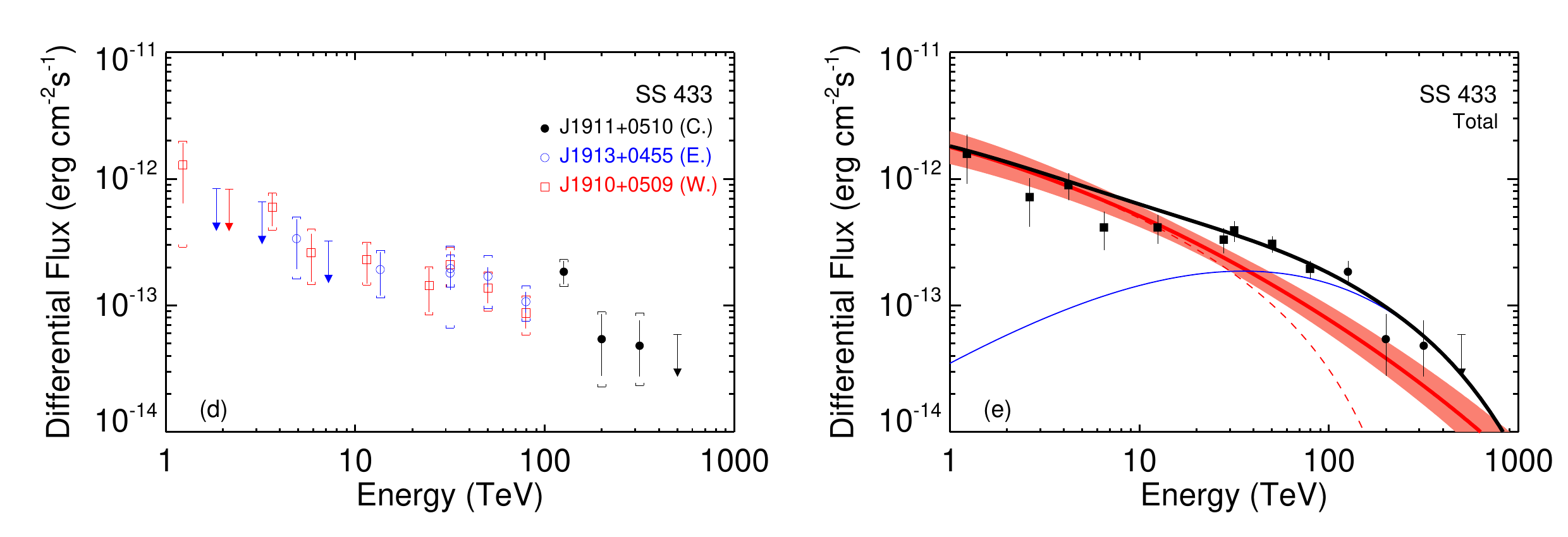}
\caption{Significance maps and spectral energy distribution (SED) of SS~433 measured by LHAASO, with surrounding sources being subtracted.
(a - c): SS~433 at energy 1-25\,TeV, 25-100\,TeV and above 100\,TeV.
In top three panels, the green cross marks the position of the BH of SS~433.
In (a), the blue diamonds shows the position of H.E.S.S. detected gamma-ray emission above 10\,TeV\cite{HESS24_SS433}.
In (a) and (b), black crosses indicate the position of resolved two point-like sources at 1-25 TeV and 25-100\,TeV.
In (c), the white contour indicates the H\,{\sc i} atomic clouds at consistent distance of SS 433 (see Supplementary Materials).
The cyan contour show the X-ray emission of the two lobes.
The green circle in (c) exhibit 68\% containment radii of the LHAASO source.
The dashed white line indicate the direction of the Galactic plane with $b=-2^\circ$.
The yellow circles show the corresponding 68\% containment radii of LHAASO PSF at the corresponding energy range.
(d) shows spectra of two point-like sources associated with the east and west lobes of SS~433 with blue circles and red { squares} respectively.  The spectrum of the central extended source is shown with the black { dots}; (e) compares the total measured spectrum (with the fluxes associated with the two lobes summed up) and the prediction of models. The red solid curve showcases the best-fit spectrum based on multiwavelength data with a single leptonic component (see SM for details), where the red band represents the $1\sigma$ uncertainty. The best-fit value of the high-energy spectral cutoff energy is about $E_{\rm e, max}=10$\,PeV. The red dashed curve shows the predicted spectrum with a conservative $E_{\rm e,max}=200\,$TeV with other parameters unchanged. The target photon fields of IC radiation include the cosmic microwave background and interstellar radiation\cite{Popescu17}. The solid blue curve shows an additional hadronic component and the solid black curve is the sum of the hadronic component and leptonic component with $E_{\rm e, max}=200\,$TeV. In (d) and (e), error bars represent the $1\sigma$ uncertainties of fluxes and bars with {downward-pointing} triangles are one-tailed 95\% upper limits of the flux.  In (d), The vertical brackets showcase the $1\sigma$ uncertainties of the flux including the systematic errors.
}
\label{fig:tsmap}
\end{figure*}

\begin{figure*}[htbp]
\centering
\includegraphics[width=1\textwidth]{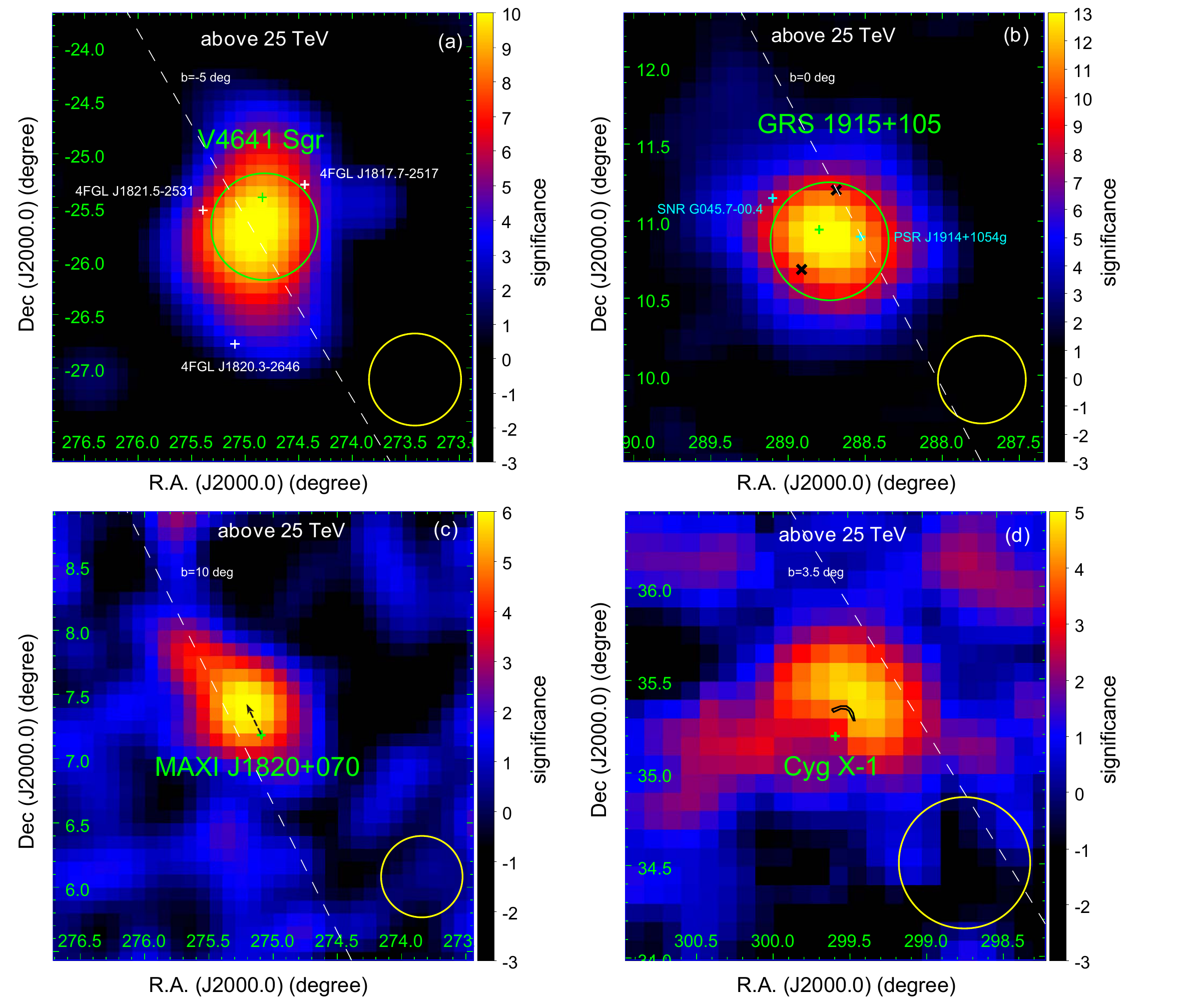}
\caption{Significance maps of four other LHAASO measured microquasars besides SS~433, (a) V4641~Sgr, (b) GRS~1915+195, (c) MAXI~J1820+070 and (d) Cygnus X-1 at above 25\,TeV,  with surrounding sources being subtracted.
In each panel, the green cross marks the position of the BH of each microquasar.
The green circles in (a) and (b) exhibit 68\% containment radii of the LHAASO sources,  whereas no green circles are shown in (c) and (d) because of the point-like nature of associated LHAASO sources.
In (a), Fermi-LAT 4FGL-DR4 GeV gamma-ray sources within 3 $\sigma$ significance region of V4641~Sgr are shown with white crosses.
In (b), other possible counterparts to the observed TeV emission are shown with cyan crosses.
The black cross represents the hot spots observed by ALMA
In panel (c), the black arrow represents the propagation direction of the receding ejecta.
The  black arc in (d) represents the bow-like radio structure inflated by the jet of Cygnus X-1\cite{2005Natur.Cygnus.X1}.
The yellow circle in each panel shows the corresponding 68\% containment radii of LHAASO PSF.
The dashed white lines indicate the direction of the Galactic plane.
}
\label{fig:tsmap2}
\end{figure*}

\begin{figure*}[htbp]
\centering
\includegraphics[width=1\textwidth]{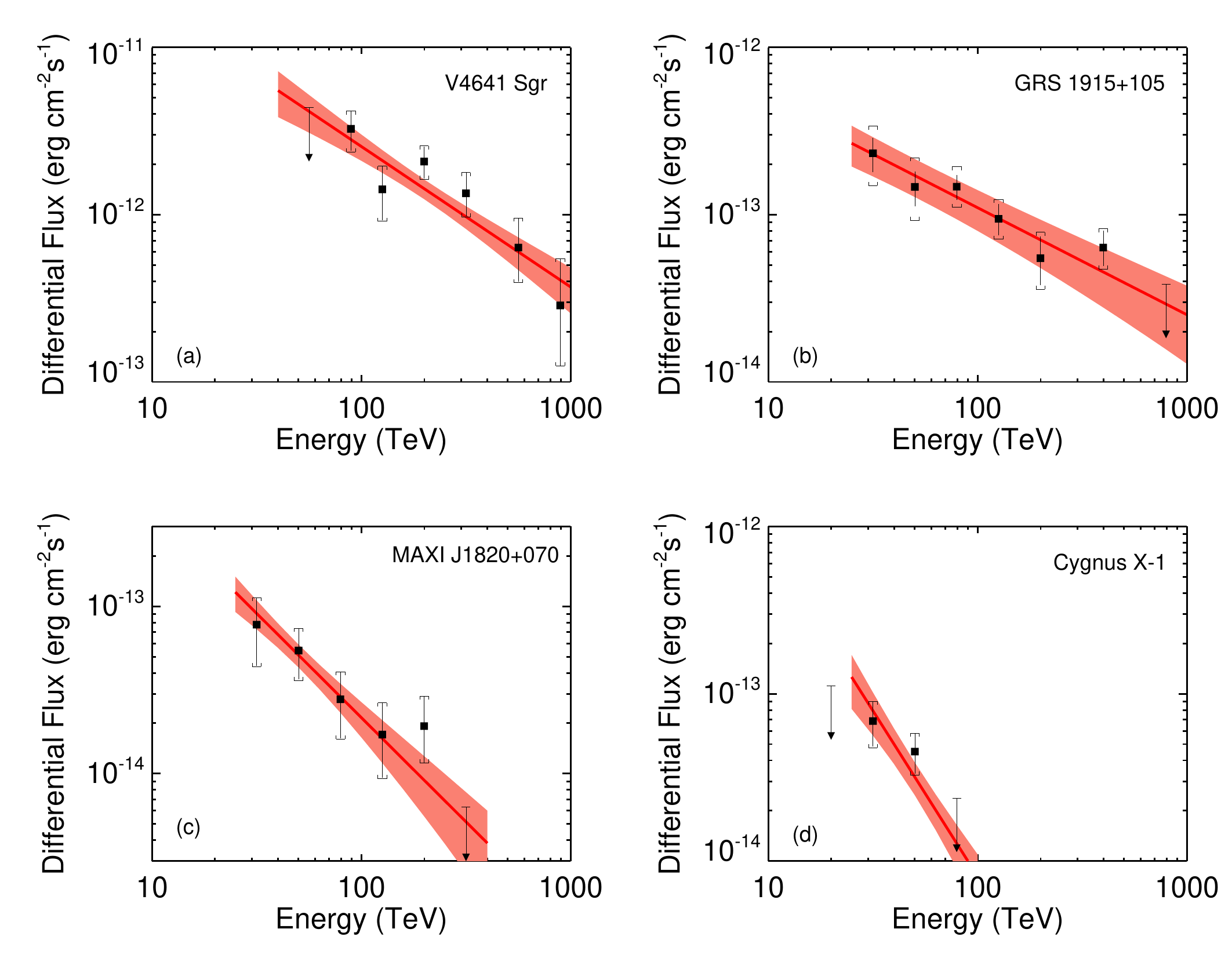}
\caption{Spectra of the LHAASO sources associated with four microquasars. (a) V4641~Sgr; (b) GRS~1915+105; (c) MAXI~J1820+070; (d) Cygnus~X-1. In each panel, error bars represent the $1\sigma$ statistical uncertainties of measured fluxes and bars with downward-pointing triangles (if present) are one-tailed 95\% upper limits of the flux.  Vertical brackets showcase the $1\sigma$ uncertainties of fluxes including the systematic errors. Red solid lines represents the best-fit spectrum with a power-law function  and shaded regions showcase uncertainties.}
\label{fig:sed}
\end{figure*}

Using the latest LHAASO dataset of photons above 25\,TeV, we detect five sources at significance levels of { $13.5\sigma$, $10.5\sigma$, $15.1\sigma$, $6.0\sigma$ and $4.4\sigma$}, respectively, associated with SS~433, V4641~Sgr, GRS~1915+105, MAXI~J1820+070 and Cygnus~X-1.
The data analysis follows the LHAASO standard pipelines using events with zenith angle $<$50$^{\circ}$ as presented in \cite{LHAAS0_catalog_2024}, and events with zenith angle between 50$^{\circ}$ and 60$^{\circ}$ were used specifically for V4641~Sgr (see Supplementary Materials, hereafter SM, for more details).
Except for Cygnus X-1, the maximum photon energies of all these sources well exceed 100\,TeV. { This implies that BH-Jet systems can efficiently accelerate particles.} There is no gamma-ray source significant over $3\sigma$ identified from the other seven microquasars. Our observational results are summarized\footnote{All the uncertainties reported in the main text are statistical errors, whereas the systematic uncertainties are sub-dominant and discussed in SM.} in Table~1. The significance map and the spectrum of SS~433 are shown in Fig.~\ref{fig:tsmap}, while those of the other four detected sources are shown in Fig.~\ref{fig:tsmap2} and Fig.~\ref{fig:sed}.

\begin{table*}[t]
    \centering
    \begin{threeparttable}
    \resizebox{\textwidth}{!}{
    \begin{tabular}{llcccccc}
    \hline\hline
      Microquasar & Distance & LHAASO Source & Significance & Photon Index & Energy Range & Extension\tnote{a} & Flux\tnote{b} \\
      ~ & ~~~(kpc) & & ($\sigma$) & ~ & (TeV) & ~ &  (Crab Unit)\\
      \hline
       SS~433 E. & \multirow{3}*{$4.9\pm 0.4$\cite{Su2018}} & J1913+0455 & 9.9\tnote{c} & $2.82\pm0.16$ & $25-100$ & \multirow{2}*{$0.73^\circ\pm 0.07^\circ$} & 0.10 \\ 
       SS~433 W. &  & J1910+0509 & 6.3\tnote{c} & $2.94\pm 0.38$ & $25-100$ &  & 0.082 \\ 
       SS~433 central &  & J1911+0510 & 8.0 & $3.96\pm 0.25$ & $100-630$ & $0.32^\circ\pm0.04^\circ$ & 0.32 \\
       V4641~Sgr & $6.2\pm 0.7$\cite{MacDonald14} & J1819-2541 & 10.5 & $2.84\pm 0.17$ & $40-1000$ & $0.33^\circ\pm 0.08^\circ$ & 2.6 \\ 
       GRS~1915+105 & $9.4\pm 0.6$\cite{Reid23} & J1915+1053 & 15.1& $2.68\pm0.13$& $25-1000$ & $0.28^\circ\pm 0.05^\circ$ & 0.11 \\ 
       MAXI~J1820+070 & $2.96\pm 0.33$\cite{Atri20} & J1821+0723 & 6.0 & $3.25\pm0.26$  & $25-400$ & $<0.28^\circ$  & 0.02 \\ 
       Cygnus~X-1 &  $2.2\pm0.2$\cite{Miller-Jones21} & J1958+3522 & 4.4 & $3.98\pm0.40$ & $25-100$ & $<0.22^\circ$\ & $<0.01$\\ 
       XTE~J1859+226 &  $4.2\pm 0.5$\cite{Shaposhnikov09} & -- & 2.7 & -- & -- & -- & $<0.02$ \\ 
       GS~2000+251 & $2.7\pm0.7$\cite{Jonker04} & -- & 2.3  & -- & -- & -- & $<0.04$ \\
       Swift J1727.8-1613 & $2.7\pm0.3$\cite{Mata24} & -- & $0.7$ & -- & -- & -- & $<0.04$ \\ 
       GRO~J0422+32 & $2.49\pm0.3$\cite{Gelino03} & -- &  0.7 & -- & -- & -- & $<0.01$ \\
       V404~Cygni & $2.39\pm0.14$\cite{Miller-Jones09} & -- & 1.5 & -- & -- & -- & $<0.03$ \\ 
       XTE~J1118+480 & $1.7\pm 0.1$\cite{Gelino06} & -- & 0.4 & -- & -- & -- & $<0.02$ \\ 
       V616~Mon & $1.06\pm 0.1$\cite{Cantrell10} & -- & 0.4 & -- & -- & -- & $<0.01$ \\
       \hline
    \end{tabular}
             }
    \caption{LHAASO's measurement of Galactic BH-jet systems in the field of view.} 
    \label{tab:list}
    \begin{tablenotes}
    \footnotesize
    \item[a] separation between two point-like sources of SS~433 below 100\,TeV; $39\%$ containment radius for SS~433 central,\\
    V4641~Sgr and GRS~1915+105; one-tailed $95\%$ confidence upper limit for the source size for MAXI~J1820+070 \\
    and Cygnus~X-1.
    \item[b] at 100\,TeV, $1\,{\rm Crab Unit}\simeq 10^{-12} \,\rm erg~cm^{-2}s^{-1}$
    \item[c] the combined detection significance for the two point-like sources is $12.3\sigma$.
    \end{tablenotes}
    \end{threeparttable}
\end{table*}

In a BH-jet system, there are several potential sites where particles can be accelerated. Termination shocks, which arise from interactions between jets and the surrounding medium, are promising particle accelerators, as  shown by nonthermal radiation observed from lobes of SS~433\cite{Bordas2010, HAWC18_SS433, HESS24_SS433}. {  Particle acceleration may also take place within the jet's outer layer of stratified velocity\cite{Rieger04} or inside the jet through internal collisions and magnetic reconnection events\cite{Bosch-Ramon09}. In addition to jet, powerful sub-relativistic wind launched from the accretion disk and the fast-rotating magnetosphere of the BH offer alternative possibilities of particle acceleration \cite{LIJ20, Gangadhara97, Rieger00}}.

{\it SS~433.} A pair of jets is launched from the central BH of SS~433.
The two jets are nearly perpendicularly to our line of sight and terminated at approximately 40 parsecs away from the BH.
We identified two point-like sources, LHAASO~J1913+0455 (RA$=288.28^\circ\pm0.04^\circ$, DEC$=4.92^\circ\pm0.04^\circ$), and LHAASO~1910+0509 (RA$=287.58^\circ\pm0.06^\circ$, DEC$=5.15^\circ\pm0.05^\circ$), in 25-100\,TeV. {  In 1-25\,TeV, we also identified two point-like sources at similar positions to those in 25-100\,TeV}. The positions of these sources are associated with the east and west X-ray lobes, respectively, and are consistent with measurements from H.E.S.S. and HAWC\cite{HESS24_SS433, HAWC18_SS433}. Above 100\,TeV, we identified an extended source, LHAASO~J1911+0510 (RA$=287.89^\circ\pm0.07^\circ$, DEC$= 5.16^\circ\pm0.07^\circ$), located $0.20^\circ$ northwest of the BH's position. The source morphology is best described with a 2-dimensional (2D) Gaussian template with $r_{39}=0.32^\circ\pm 0.04^\circ$, {  covering the central BH}. The energy-dependent morphology of SS~433 suggests a different origin for the emission above 100\,TeV compared to that below 25\,TeV, or at least an additional contribution from particles other than those responsible for two lobes detected at energy below tens of TeV.
Notably, the source coincides spatially with a giant atomic cloud of  $1.0\times 10^5\,M_\odot$ { (see SM)}, as shown by the white contours in Fig.~\ref{fig:tsmap}c. This spatial coincidence may suggest a hadronic origin of the UHE emission from SS~433, via the proton-proton collision between hydrogen nuclei in the cloud and high-energy protons arriving at the cloud. Indeed, we found indications of multiple components in the emission above 100\,TeV, consisting of two point-like sources associated with the lobes and one extended source associated with the atomic cloud (see Supplementary Materials).
We performed a phenomenological {  fitting to the multiwavelength flux} of the two point-like sources with a {  one-zone} leptonic model, {  assuming a constant injection luminosity of relativistic electrons. The influence of radiative cooling on the electron spectrum, which leads to a softening at high energies, is taken into account.}
As shown in Fig.~\ref{fig:tsmap}e, our data up to 30\,TeV can be well explained by the IC radiation of electrons, consistent with the suggestion by the H.E.S.S. experiment. However, the data around 100\,TeV cannot be reproduced even with an extremely high (and probably unrealistic) spectral cutoff energy $E_{\rm e, max}\approx 10\,$PeV, which is the best-fit value from the fitting, due to the suppression by the Klein-Nishina (KN) effect. Using the most conservative value of $E_{\rm e, max}=200\,$TeV as suggested by H.E.S.S measurements would entail a more significant contribution from an additional spectral component.  {Although a secondary leptonic component remains possible,  we focus here on a hadronic scenario motivated by the proximity of the nearby atomic cloud. In this framework, we assume that an additional population of protons is injected into the surrounding medium by the microquasar, diffuses over larger distances, and produces high-energy photons upon interacting with the cloud.} We found that the data can be satisfactorily explained under this picture. The model suggests that SS~433 is continuously injecting PeV protons at a power of $\sim 10^{38}\,$erg/s into the surrounding medium (see Supplementary Materials for more details). These energetic protons could be produced either close to the BH at the centre of the system or at the two lobes (i.e., jet terminations). Deeper observations in the UHE gamma-ray band, as well as those in the X-ray and radio bands, are crucial for unraveling { the origin of the additional spectral component}.

{\it V4641~Sgr.} An extended source, LHAASO~J1819-2541 was discovered in the vicinity of V4641~Sgr.
The centre of the source (RA$=274.82^\circ\pm 0.09^\circ$, DEC$=-25.60^\circ\pm 0.11^\circ$) is offset from V4641~Sgr by about $0.19^\circ$.
There are no other reasonable astrophysical counterparts with better spatial association to the LHAASO source, making V4641~Sgr the best candidate origin of LHAASO~J1819-2541.
The measured spectrum can be described by a power-law function with an index of -2.84$\pm$0.17, {  which is softer than the spectrum obtained by HAWC within a lower energy range of $10-200$\,TeV\cite{HAWC24} (see Supplementary Fig.~12 in SM)}. The highest energy of photons detected from this source is extended up to 0.8\,PeV with LHAASO's observation. IC radiation of electrons cannot reproduce such a hard spectrum continuing to almost 1\,PeV due to the KN effect, unless a very hard injection spectrum is assumed or the IC radiation dominates the cooling even in the KN regime {(which would require a very weak magnetic field of $B<0.5\mu$G for 1\,PeV electrons)}. On the other hand, if the emission is of hadronic origin, a more reasonable spectrum of protons can be obtained. In this case, since the spectrum does not show a clear cutoff feature, V4641~Sgr would be a so-called super-PeVatron that energizes protons up to energies of at least $\sim 10$\,PeV.

{\it GRS~1915+105.} As the first known Galactic object ejecting matters with relativistic motion\cite{Mirabel94},
GRS~1915+105 is the famous archetypal microquasar. We discovered an extended source, LHAASO~J1915+1053 (RA$=288.74^\circ\pm 0.05^\circ$, DEC$=10.88^\circ\pm 0.05^\circ$), located at $0.1^\circ$ southwest of GRS~1915+105. There are other potential high-energy particle accelerators, such as PSR~J1914+1054g and SNR~G045.7-00.4, around the LHAASO source, but {  GRS~1915+105 is most likely the counterpart} (see Supplementary Materials for detailed discussion).
Interestingly, ALMA has discovered two hotspots both at $0.28^\circ$ (corresponding to about 50\,pc) away from GRS~1915+105 in the opposite directions of the BH\cite{Tetarenko18}, which may be interpreted as two lobes driven by jets. {  Recent observation by MeerKAT supports this picture\cite{Motta2025}.}  Indeed, the angular separation between the two ALMA hotspots is comparable to the extension of the LHAASO source with $r_{39}=0.28^\circ$. It could be due to the large distance of GRS~1915+105 (i.e., 9.4\,kpc\cite{Reid23}), so that LHAASO does not resolve their emission.
Alternatively, the extended nature of the source and the offset from the BH may be ascribed to the spatial distribution of the surrounding gas, if the emission from GRS~1915+195 is dominated by hadronic processes.

{\it MAXI~J1820+070 and Cygnus~X-1.} Two point-like sources, LHAASO~J1821+0723 (RA$=275.22^\circ\pm 0.05^\circ$, DEC$=7.39^\circ\pm 0.06^\circ$) and LHAASO~J1957+3517 (RA$=299.47^\circ\pm 0.13^\circ$, DEC$=35.37^\circ\pm 0.06^\circ$), are discovered in spatial association to MAXI~J1820+070 and Cygnus~X-1, respectively. MAXI~J1820+070 was discovered in 2018 during its X-ray outburst\cite{Tucker18}. A pair of bipolar relativistic radio ejecta was launched from the central black hole during the outburst\cite{Bright20}. Although the LHAASO source is detected at $0.27^\circ$ northeast of the BH {  (corresponding to a $3.5\sigma$ offset)}, it is in alignment with the propagation direction of the receding ejecta. There is also a {  moderate offset of $0.19^\circ$ (corresponding to $1.6\sigma$)} between Cygnus~X-1 and the associated LHAASO source . The offset follows the same direction as the radio-emitting bow-like structure (shown as the cyan arc in Fig.~\ref{fig:tsmap2}d) that appears to be inflated by the jet launched from the central BH\cite{2005Natur.Cygnus.X1}.
Extension of these two LHAASO sources are not evident, and in turn we obtain upper limits of $0.28^\circ$ and $0.22^\circ$ for the sizes of J1821+0723 and J1957+3517 respectively.
While we do not expect variability of emission from extended sources, the flux of a point-like source may in principle vary with time. We therefore check their temporal behaviours, but find no evidence of variability. Future continuous observation of LHAASO is crucial for further clarification of the morphological, spectral and temporal properties of these two sources.

\section{DISCUSSIONS AND CONCLUSIONS}\label{sec3}
{Among the five microquasars with TeV-PeV emission, SS~433 and Cygnus~X-1 exhibit persistent BH activity, whereas V4641~Sgr and GRS~1915+105 have shown frequent radio or X-ray outbursts in recent years \cite{Shaw22, Mendez2022}. On the other hand, most of seven microquasars without significant LHAASO detection presented only one historical outburst leading to their discoveries (with the exception of V404~Cyg, which underwent outbursts in 2015\cite{Bright20}). {While the timescales for any of the relevant radiation mechanisms are too long to directly link the activity in the last few decades to the UHE emission, the observed pattern implies that the UHE emission serves as a proxy for} the long-term activity status of the BH binaries. In that case, the subset of undetected microquasars with frequent flaring activity would be the most likely candidate for further UHE discoveries. Conversely, if the apparent connection is coincidental, we would expect UHE emission associated with many currently quiescent BH binaries that remain unidentified. Future joint observations of LHAASO and X-ray/radio instruments may elucidate this.}

Detection of UHE gamma-ray { emission} from multiple microquasars establishes that BH-jet systems are potent particle accelerators.
{The energy-dependent morphology of SS~433 and the spatial coincidence between its UHE emission and a nearby atomic cloud suggests a scenario of the coexisting leptonic and hadronic emission components, supporting its capacity to accelerate protons to the ``knee'' of the CR spectrum. Meanwhile, the spectrum of V4641~Sgr has been measured up to 800\,TeV, further positioning microquasars as potential super-PeVatrons.}

A crucial follow-up question is whether these Galactic BH-jet systems are responsible for the origin of CRs above PeV energies. Extrapolating from SS~433’s PeV proton luminosity inferred from our model ($L_{p, \rm PeV}\lesssim 10^{38}\,\rm erg~s^{-1}$), we estimate that the Milky Way’s microquasar population could collectively inject PeV protons at $f_{\mu \rm Q}L_{p,\rm PeV}=10^{39}\,\rm erg~s^{-1}$, where $f_{\mu \rm Q}\sim 10$ accounts for the scaling factor in source number and power of the entire population in the Milky Way. This matches the observed PeV proton flux on Earth within the leaky-box model framework (see Supplementary Materials for more details). The acceleration mechanisms and acceleration sites in these BH-jet systems, on the other hand, have not been clearly identified based on our current observations. In principle, particle acceleration may be processed at different scales and environments, ranging from the vicinity of BHs to the endpoints of jets { (such as re-collimation shocks and termination shocks\cite{Heinz2002, Bordas09})}. Future observations with higher statistics and dedicated analyses of each microquasar, along with multiwavelength observations, can conduct more detailed spectral, morphological and temporal measurements. This will facilitate our understanding of physical processes in these Galactic stellar-mass BH-jet systems, and provide insights into UHECR acceleration in their more powerful siblings at distant universe harboring supermassive BHs, such as radio galaxies and blazars.

\clearpage

\appendix

\section{LHAASO DATA\label{app1}}
In this work we utilize the data acquired by LHAASO-KM2A 1/2 array from December 26, 2019 to November 30, 2020 (with a live time of 290 days), by the 3/4 array from December 1, 2020 to July 19, 2021 (with a live time of 216 days), and by the full array from July 20, 2021 to December 31, 2024 (with a live time of 1228  days).
The performance of  LHAASO-KM2A have been studied in detail employing the Monte Carlo simulations \cite{2024RDTM..tmp...28C}, and calibrated using the measurements of Crab Nebula as a standard candle \cite{2021ChPhC..45b5002A} (see also SM).  The data quality control system and the long-term performance monitoring about the  LHAASO-KM2A data can be found in ref.\cite{2024arXiv240511826C}. The pipeline of LHAASO-KM2A data analysis presented in ref.\cite{2021ChPhC..45b5002A} is directly adopted in this analysis.
Considering the energy resolution and statistics, one decade of energy is divided into five bins with a bin width of $\log_{10}E=0.2$.  The sky in celestial coordinates is divided into grids with size of $0.1^{\circ} \times 0.1^{\circ}$  and filled with detected events according to their reconstructed arrival directions for each energy bin.
The background map is estimated by the direct integration method\cite{Fleysher2004ApJ}.

LHAASO-WCDA data adopted in this analysis covers from March 8, 2021 to May 31, 2024, with a total live time of 1092\,days.
Events with zenith angle larger than 50\,$^{\circ}$ or RMDS larger than 20 meters are not included to keep good event quality, where RMDS is the root of mean distance square of the top 10 hottest detector units which represents the compactness of lateral distribution of air showers.
The number of fired detector units, $N_{\rm hit}$, is selected as the estimator of primary energy.
The events are divided into six segments with $N_{\rm hit}$ value of 60-100, 100-200, 200-300, 300-500, 500-800 and 800-2000. We do not use events of larger $N_{\rm hit}$, because when $N_{\rm hit}$ is {  larger than 2000}, the instrument is close to saturation and the energy resolution becomes poor.
The {  direct integration method} is adopted to estimate the cosmic ray background.

With spectrum model of LHAASO sources assumed as a simple power-law and spatial model as Gaussian extension, an iteration process based on multi-dimensional maximum likelihood is performed to derive the spectrum, position and extension.
The iteration process adds one source at a time to the fitting as long as the TS of N+1 sources is greater than that of N sources by more than 25.
The dust column density measured by PLANCK is introduced as the spatial template of diffuse Galactic gamma-ray emission.
For more details, see the Supplementary Materials.

\section{SUPPLEMENTARY DATA}
Supplementary data are available at NSR website.

\section{ACKNOWLEDGEMENTS}
We would like to thank all staff members who work at the LHAASO site above 4400 meter above the sea level year round to maintain the detector and keep the water recycling system, electricity power supply and other components of the experiment operating smoothly. We are grateful to Chengdu Management Committee of Tianfu New Area for the constant financial support for research with LHAASO data. We appreciate the computing and data service support provided by the National High Energy Physics Data Center for the data analysis in this paper.

\section{FUNDING}
This research work is supported by the following grants: The National Natural Science Foundation of China No.12393851, No.12393852, No.12393853, No.12393854, No.12333006, No.12273038, No.12205314, No.12105301, No.12305120, No.12261160362, No.12105294, No.U1931201, No.12375107, No.12173039, the Department of Science and Technology of Sichuan Province, China No.24NSFSC2319, Project for Young Scientists in Basic Research of Chinese Academy of Sciences No.YSBR-061,
and in Thailand by the National Science and Technology Development Agency (NSTDA) and the National Research Council of Thailand (NRCT) under the High-Potential Research Team Grant Program
(N42A650868).

\section{AUTHOR CONTRIBUTIONS}
Z. Cao initiated the project. R.Y. Liu and J. Li designed the study and led the writing of the paper. Y.H. Yu, J. Li, R.Z. Yang and S.C. Hu analysed the data of the SS~433. S.Z. Chen analysed the data of V4641~Sgr. K. Wang analysed the data of GRS~1915+105, MAXI~J1820+070, and seven undetected microquasars. C. Li analysed the data of Cygnus X-1. S.Q. Xi did the cross check of the LHAASO data analyses. J. Li and S. Yang analysed  H\,{\sc i} data around SS~433. R.Y. Liu led the interpretation of the data. F. Aharonian and H. Feng provided important comments on the manuscript. All the authors discussed and edited the manuscript.

\bibliographystyle{nsr}
\bibliography{ms}

\newpage
\onecolumn
\begin{center}
Zhen Cao$^{1,2,3}$,
F. Aharonian$^{3,4,5,6}$,
Y.X. Bai$^{1,3}$,
Y.W. Bao$^{7}$,
D. Bastieri$^{8}$,
X.J. Bi$^{1,2,3}$,
Y.J. Bi$^{1,3}$,
W. Bian$^{7}$,
A.V. Bukevich$^{9}$,
C.M. Cai$^{10}$
% ...
(The LHAASO Collaboration)

$^{1}$ Key Laboratory of Particle Astrophysics \& Experimental Physics Division \& Computing Center, Institute of High Energy Physics, Chinese Academy of Sciences, 100049 Beijing, China\\
$^{2}$ University of Chinese Academy of Sciences, 100049 Beijing, China\\
$^{3}$ TIANFU Cosmic Ray Research Center, Chengdu, Sichuan,  China\\
$^{4}$ University of Science and Technology of China, 230026 Hefei, Anhui, China\\
$^{5}$ Yerevan State University, 1 Alek Manukyan Street, Yerevan 0025, Armeni a\\
$^{6}$ Max-Planck-Institut for Nuclear Physics, P.O. Box 103980, 69029  Heidelberg, Germany\\
$^{7}$ Tsung-Dao Lee Institute \& School of Physics and Astronomy, Shanghai Jiao Tong University, 200240 Shanghai, China\\
$^{8}$ Center for Astrophysics, Guangzhou University, 510006 Guangzhou, Guangdong, China\\
$^{9}$ Institute for Nuclear Research of Russian Academy of Sciences, 117312 Moscow, Russia\\
$^{10}$ School of Physical Science and Technology \&  School of Information Science and Technology, Southwest Jiaotong University, 610031 Chengdu, Sichuan, China\\
% ...

\end{center}

\includepdf[pages=-, fitpaper=true, offset=20mm -20mm]{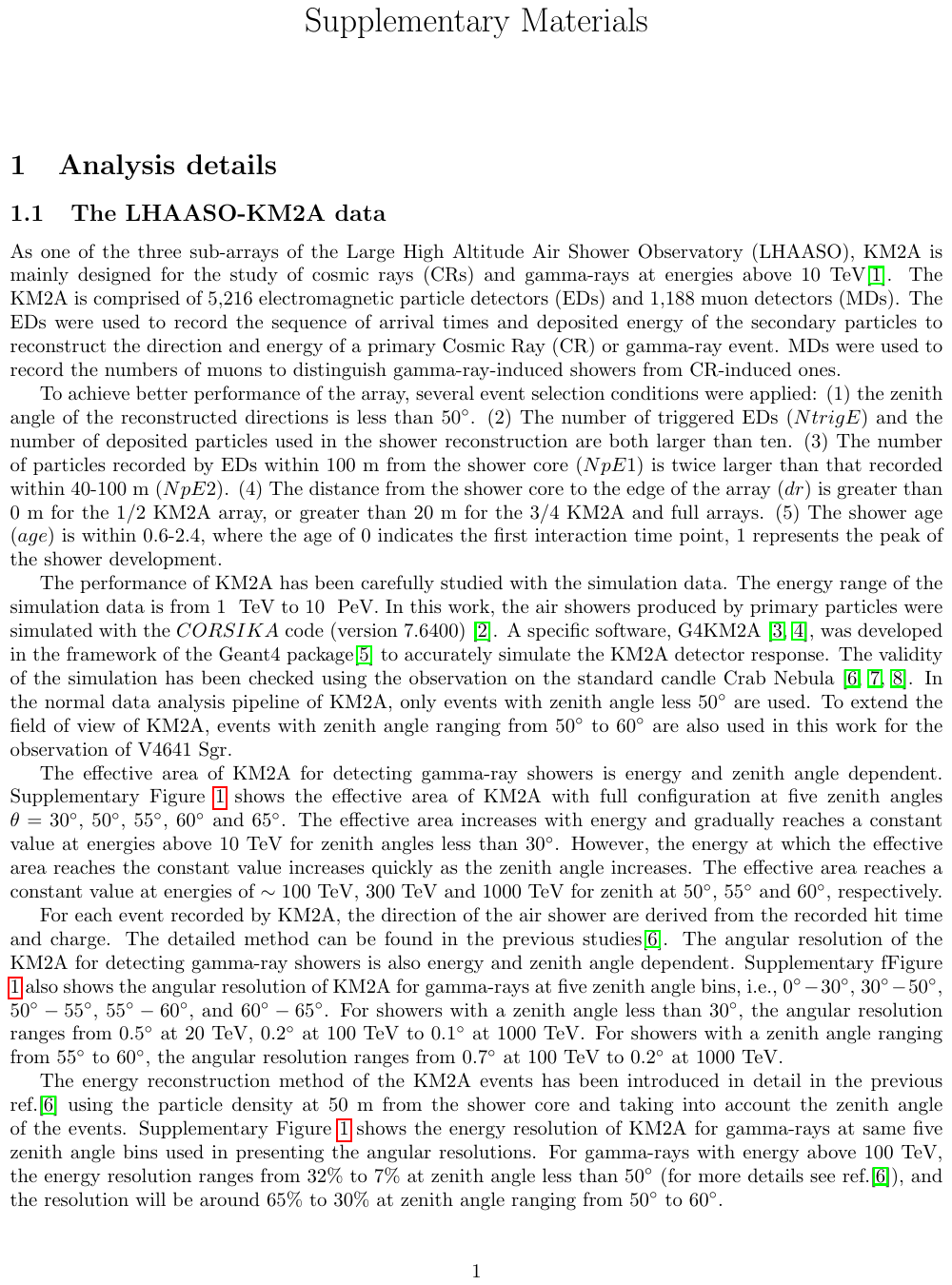}

\end{document}